%% file: LitReview.tex
\documentclass{llncs}
\usepackage[a4paper,margin=1in]{geometry}

\usepackage[T1]{fontenc}

\usepackage{xcolor}
\usepackage{amsfonts}

\usepackage{amsmath}
\usepackage{multirow}

\usepackage{algorithmicx}
\usepackage{algpseudocode}

\usepackage{lscape}

\usepackage{array}
\newcolumntype{H}{>{\setbox0=\hbox\bgroup}c<{\egroup}@{}}

\newcommand\todo[1]{\textcolor{red}{#1}}

\usepackage{pifont}
\newcommand{\cmark}{\ding{51}}%
\newcommand{\xmark}{\ding{55}}%

\newcommand\bigO[1]{\mathcal{O}(#1)}
\newcommand\tO[1]{\widetilde{\mathcal{O}}(#1)}
\makeatletter
\newcommand{\doublewidetilde}[1]{{%
  \mathpalette\double@widetilde{#1}%
}}
\newcommand{\double@widetilde}[2]{%
  \sbox\z@{$\m@th#1\widetilde{#2}$}%
  \ht\z@=.9\ht\z@
  \widetilde{\box\z@}%
}
\makeatother
\newcommand{\tto}[1]{\doublewidetilde{o}(#1)}

\usepackage{xspace}

\usepackage{comment}

\newcommand{\etal}{\emph{et al.}\xspace}
\newcommand{\ie}{\emph{i.e.}\xspace}
\newcommand{\eg}{\emph{e.g.}\xspace}

\usepackage{pgf}
\usepackage{tikz}
\usetikzlibrary{arrows,automata}
\usepackage[latin1]{inputenc}
\usepackage{verbatim}
\usepackage{graphicx}

\begin{document}
\title{Dynamic Shortest Path and Transitive Closure Algorithms:\\A Survey}
\author{Daniel P. Martin}
\institute{School of Mathematics, University of Bristol, Bristol, BS8 1TW, UK,\\ and the Heilbronn Institute for Mathematical Research, Bristol, UK\\\email{dan.martin@bristol.ac.uk}}
\maketitle
\begin{abstract}
Algorithms which compute properties over graphs have always been of interest in computer science,
with some of the fundamental algorithms, such as Dijkstra's algorithm, dating back to the 50s.
Since the 70s there has been interest in computing over graphs which are constantly
changing, in a way which is more efficient than simply recomputing after each time the graph
changes.

In this paper we provide a survey of both the foundational, and the state of the art, algorithms
which solve either shortest path or transitive closure problems in either fully or partially
dynamic graphs. We balance this with the known conditional lowerbounds.
\end{abstract}

\section{Introduction}

Graphs are one of the most fundamental and well studied data structures within computer science.
Many problems of interest can be phrased in terms of graphs. In this work we focus
on two problems over graphs; connectivity and shortest paths;

\noindent\emph{Connectivity}: Given two vertices within a graph, does there exist a path between the
two vertices? If the graph is directed, asking if there is a path between vertex $u$ and vertex $v$
is called \emph{transitive closure}. If a path is required both from $u$ to $v$ and from $v$
to $u$, the question is asking if $u$ and $v$ are in the same \emph{strongly connected component}.
If the graph is undirected, two vertices being connected and being in the same strongly connected component
are equivalent.

\noindent\emph{Shortest Path}: Given two vertices $u, v$ in a graph what is the shortest path from
$u$ to $v$? A slight variant of the problem only requires the length of the shortest path to be returned
and not the component itself. The problems described above are the exact variant; there is also an approximate
variant where given $\alpha>1$ a path of length $\alpha\cdot\Delta$ must be returned, where $\Delta$ is the length
of the shortest path.

For both of these problems there are a few variants. In the single source variant, the queries will only be
asked from a fixed start vertex. A similar variant can be considered for a single sink. In the
$s$-$t$ variant, the only query asked will be between the vertices $s$ and $t$.

These two problems occur in a host of areas, including; databases~\cite{PoDS:Yannakakis90}, compilers~\cite{IST:Reps98}
and VLSI design~\cite{MISC:Chen96}. For example, in bioinformatics, shortest path algorithms have been used to identify
genes associated with colorectal cancer~\cite{MISC:LHLC+12}.
Another example, is that if you represent the possible states of a
Rubik's cube as vertices in the graph, where an edge corresponds to a single twist, then a shortest path algorithm
will give an efficient way to solve any cube~\cite{PROJ:Kaur15}.

The problems defined above are given in a static setting; when the graph is defined, it will remain fixed
for the entirety of the graph's lifetime. However, in practice there may be times when the graph will change
over the course of the algorithm. In these situations it is desirable not to have to start from scratch --
dynamic graph algorithms are for these situations. There are three variants of dynamic graph algorithms;
\emph{incremental} algorithms which only allow edges to be inserted and not deleted, \emph{decremental}
algorithms which only allow edges to be deleted and \emph{fully dynamic} algorithms which allow edges
to both be inserted and deleted.\footnote{In this work we will not consider the case where vertices can be inserted or deleted.}
Some algorithms on weighted graphs, also allow the edge weights to be changed.

Fully dynamic algorithms have a multitude of clear uses, as in many applications the graph will naturally
change over time. Incremental algorithms are
useful in scenarios where links appear but never disappear. For example, representing the graph of all people
on a social network, where the edges are if two people have ever communicated. The road network tends to be
constantly expanding, with it being extremely rare for a road to be permanently destroyed. 
The use of decremental algorithms, is slightly more subtle but they are of important theoretical interest.
This is because they can be used as a building block in a variety of fully dynamic algorithms~\cite{JOC:HenKin10,FOCS:Bernstein09,STOC:AbrCheGav12,LIPI:AbrCheTal14,JOC:RodZwi08}.
Lots of decremental algorithms can be converted into incremental algorithms, while maintaining the same time
complexities, such as~\cite{STOC:BerChe16}. Decremental algorithms have even been used to give improved static
graph algorithms~\cite{STOC:Madry10}.

The are other ways that a graph can be considered to be changing dynamically. For example, Albers and
Schraink~\cite{ARXIV:AlbSchr17} consider graph colouring where the vertices of the graph arrive one
at a time. We will not focus on this, or other models, in this survey but limit ourselves to the case
where edges are added and removed.

Dynamic algorithms can trivially be constructed from the classic approach. If the
algorithm does nothing when an edge is updated, it will have to calculate the answer to any query from scratch.
This gives a $\bigO{1}$ update time and $\tO{m+n}$ query time algorithm for all pairs shortest path. However, if after each change the
static algorithms are rerun, this results in a $\tO{n\cdot m+n^2}$ update time and $\bigO{1}$ query time all pairs shortest path algorithm.
The goal of research into dynamic graph algorithms is to do better than these trivial bounds. For incremental
connectivity, a set union algorithm~\cite{JACM:Tarjan75} can be used to give a query and update time of $\bigO{\alpha(n)}$,
where $\alpha$ is related to the inverse-Ackerman function and thus very slow growing.
In this work
we focus mainly on algorithms with a constant query time. However, there are a host of algorithms which do
better than the trivial case without this property~\cite{FOCS:HenKin95,FOCS:King99,FOCS:RodZwi04,FOCS:Bernstein09}. For example, Roddity and Zwick~\cite{FOCS:RodZwi04}
give a fully dynamic  $(1+\epsilon)$-approximation algorithm for dynamic all pairs shortest paths with
$\tO{\frac{m\cdot n}{t}}$ update time and $\bigO{t}$ query time. Hence, this algorithm provides a trade-off
between update and query time.

Even and Shiloach~\cite{JACM:EveShi81} designed the first
decremental connectivity algorithm, which is more efficient than simply recomputing from scratch after
the deletion of an edge. A similar result was independently discovered by Dinitz~\cite{TCS:Dinitz06}.
The algorithm of Even and Shiloach will be the starting point for our survey, since it has had a
great deal of influence in the literature, resulting in several variants and multiple algorithms being
built on top of it~\cite{FOCS:HenKin95,FOCS:King99,SODA:BerRod11,STOC:BerChe16,FOCS:HenKriNan14,STOC:HenKriNan14}.
There have also been a host of algorithms which use techniques disjoint from those given by Even and
Shiloach~\cite{SODA:BerRod11,STOC:DemIta03,TALG:Lac13,SODA:Roditty13}. In contrast to the efficient algorithms
to solve these problems, there have been several lower bounds given in the literature~\cite{STOC:Patrascu10,FOCS:AbbWill14,SODA:KopPerPor16,STOC:HKNS15}.
Table~\ref{tbl:upper} summarises a large proportion of the results in the field, while the text focuses on a
smaller proportion, to go into more detail on.

This paper focuses on surveying the theoretical aspects of dynamic graph algorithms for connectivity and shortest
path. However, there has been previous work that compares these algorithms in practical scenarios~\cite{ESA:DFIT06,JEA:FINP98,ESA:FMNP+98,IWEA:DFMN00,JEA:FMNZ01,TALG:DemIta06,JEA:AlbCatIta97,JEA:IKRT01,JEA:Zaroliagis02}.
For example,
Frigioni \etal~\cite{JEA:FMNZ01} implement and compare dynamic transitive closure algorithms, while
Demetrescu and Italiano~\cite{TALG:DemIta06} implement and compare dynamic all pairs shortest path algorithms.

In this paper, we have limited ourselves to the aforementioned two problems, on general graphs.
However, there has been lots of work which considers other dynamic problems.
These included how to embed a dynamically changing graph into the plane~\cite{LIPI:HolRot15},
topological ordering~\cite{SODA:BenFinGil09,TALG:HKMS+12}, matchings~\cite{SODA:Sankowski07}
and min-cut~\cite{STOC:Thorup01}.
It is still an open question as to if fully dynamic max flow can be done faster than recomputation.
There has also been work which gives more efficient solutions when restrictions are made to the graph, such as if the graph is
planar~\cite{FOCS:FakRao01,STOC:AbrCheGav12,SODA:ACDG+16,FOCS:AbbDah16}, has bounded degree~\cite{AI:Yellin93} or is directed acyclic~\cite{IPL:Italiano88,JACM:DemIta05}.

There are two types of algorithm which can be considered; randomised or deterministic, both of which will be discussed in this paper. In general, randomised
algorithms tend to be more efficient than deterministic algorithms. However, in the dynamic graph scenario, the randomised algorithms come with a large
disadvantage. The randomised algorithms tend to assume that the adversary\footnote{Here we define an adversary as a user of the algorithm who is trying to
get the algorithm to run in the worst possible time.} is both oblivious to the randomness used by the algorithm and non-adaptive in
their queries. The non-adaptivity is required because otherwise it is possible that queries would allow the adversary learn about the randomness used by the
algorithm and then act upon this information. Without this assumption the algorithms tend to lose their benefit. Using the nice example from~\cite{STOC:BerChe16};
If the goal is to maintain a set of approximately $\sqrt{n}$ vertices called centers such that every vertex is at most $\sqrt{n}$ away from a center, in the
static case this can be achieved by simply choosing $\sqrt{n}\cdot\log{n}$ centers uniformly at random. In the deterministic setting this does not work because
the adversary can simply choose to disconnect all the centers (while leaving the rest of the graph intact), by deleting a suitable subset of edges.
In the randomised setting with oblivious adversaries the randomised solution once again holds. Hence, one
of the desirable goals for dynamic algorithms, is to construct deterministic algorithms with the same time complexity as their randomised counterparts.

\subsection{Related Work}
Recently Madkour \etal~\cite{ArXiv:MARR+17} published a survey on shortest path algorithms and fitted algorithms into a framework to aid relating algorithms. They give
a section providing an overview of dynamic algorithms. However, due to the wide subject area, they only touch on dynamic algorithms, while we can give considerably
more detail.

Demetrescu and Italiano~\cite{JoDA:DemIta06} provided a survey of dynamic shortest path and transitive closure algorithms.
The authors abstract out some of the combinatorial, algorithmic and data structure techniques used in the literature and then
present a subset of the known algorithms within this unifying framework.
However, there has been a host of work published since the survey was published over a decade ago.

\subsection{Outline}
In Section~\ref{sec:prelim} we give some preliminaries. Section~\ref{sec:estree} gives the algorithm of Even and Shiloach~\cite{JACM:EveShi81} and some variants. Section~\ref{sec:usinges} gives algorithms which are built upon the algorithm of Even and Shiloach, while Section~\ref{sec:noes} gives algorithms built from other techniques. In Section~\ref{sec:limits} we describe the known lower bounds on these problems, which reduce to other well studied problems. We conclude in Section~\ref{sec:open} with some open problems.

\section{Preliminaries}\label{sec:prelim}
In this section we give the notation that will be used throughout
this paper, along with the definitions of problems which the given
algorithms will try to solve.

\subsection{Notation}
Let $G=(V,E)$ be a (possibly directed) graph. We denote $n=|V|$ and
$m=|E|$. If the graph is weighted, we use the weight function $\ell(u,v)$
for $u,v\in V$ to give the weight, of the edge, between $u$ and $v$. Given $U\subseteq V$,
we define $G[U]=(U,E')$ to be the graph with vertices from $U$ and $E'= E\cap U^2$,
\ie all edges from $G$ that start and end in $U$.

A series of vertices $p=\langle v_0,\ldots,v_k\rangle$ is said to be a \emph{path} between $u$ and $v$,
of length $k$ if $v_0=u,v_k=v$ and $(v_i,v_{i+1})\in E\ \forall 0\leq i<k$. The weight
of a path $\ell(p)$ is defined as the sum of the weight of all the edges in the path.
A path $p$ is called a \emph{shortest path} between $u$ and $v$ if $\ell(p)\leq\ell(p')$ for
all other paths $p'$ between $u$ and $v$.

Given three variables $n_1,n_2,n_3$ we say a function $f(n_1,n_2,n_3)=\tto{n_1^{c_1}\cdot n_2^{c_2}\cdot n_3^{c_3}}$
if there exists a constant $\epsilon>0$ such that $f(n_1,n_2,n_3)=\bigO{n_1^{c_1-\epsilon}\cdot n_2^{c_2}\cdot n_3^{c_3}+n_1^{c_1-\epsilon}\cdot n_2^{c_2-\epsilon}\cdot n_3^{c_3}+n_1^{c_1}\cdot n_2^{c_2}\cdot n_3^{c_3-\epsilon}}$. It can be defined analogously for an arbitrary number
of parameters.

\subsection{Definitions}

Given a graph $G$, a \emph{dynamic} algorithm is one which allows changes
to the graph intermixed with the queries, while a \emph{static} algorithm
does not allow changes to the graph. An algorithm which only allows edge
deletion is called \emph{decremental}, an algorithm which only allows
edge insertion is called \emph{incremental}, while one which allows both
is called \emph{fully dynamic}.

The type of queries allowed determine which problem the algorithm can solve.
In this work we consider two types of problem; shortest path and transitive closure.
We define both in the static setting but the algorithms are trivially extended
to the dynamic setting by allowing the relevant changes to the graph.

\begin{definition}[Transitive Closure Problem]
 Given a directed graph $G=(V,E)$, the transitive closure problems require an algorithm to answer
 questions of the form:
 \begin{description}
  \item[Transitive Closure] ``Given $u,v\in V$, is there a path from $u$ to $v$?''
  \item[Single Source Transitive Closure] ``Given $u\in V$, is there a path from the fixed vertex $s$ to $u$?''
  \item[Single Sink Transitive Closure] ``Given $u\in V$, is there a path from $u$ to the fixed vertex $t$?''
   \item[$s$-$t$ Transitive Closure] ``Is there a path from the fixed vertex $s$ to the fixed vertex $t$?''
 \end{description}
\end{definition}

\paragraph{Note:} When the graph is not directed, transitive closure is referred to as connectivity.

\begin{definition}[Shortest Path Problem]
 Given a directed graph $G=(V,E)$, the shortest path problems require an algorithm to answer
 questions of the form:
 \begin{description}
  \item[All Pairs Shortest Paths] ``Given $u,v\in V$, return the shortest path from $u$ to $v$''
  \item[Single Source Shortest Paths] ``Given $u\in V$, return the shortest path from the fixed vertex $s$ to $u$''
  \item[Single Sink Shortest Paths] ``Given $u\in V$, return the shortest path from $u$ to the fixed vertex $t$''
   \item[$s$-$t$ Shortest Paths] ``Return the shortest path from the fixed vertex $s$ to the fixed vertex $t$''
 \end{description}
\end{definition}

\paragraph{Note:}  A variation of the shortest path problem only requires the algorithm to return the length of the shortest path,
 instead of the actual path.

\begin{definition}[Strongly Connected Component Problem]
 Given a directed graph $G=(V,E)$, the strongly connected component problem requires an algorithm to answer
 the question:

 ``Given $u,v\in v$ are $u$ and $v$ in the same strongly connected component?''
 
\noindent
 where a set of vertices $v_1,\ldots, v_k$ are said to form a strongly connected component $C$ if for all
 $v_i,v_j\in C$ there is a path from $v_i$ to $v_j$ and a path from $v_j$ to $v_i$.
 
Similar source and sink variants can also be defined.
\end{definition}

\paragraph{Note:} If the graph is undirected, two vertices being connected is equivalent to them being in the same strongly
 connected component.

\input{spg}

Several of the algorithms need the notion of a shortest path graph from source $s$. This is defined below
and an example is given in Figure~\ref{fig:SPGexample}.

\begin{definition}[Shortest Path Graph]
Given a graph $G$ and a source $s$, the shortest path graph $H_s$ is defined as the union of all shortest
paths in $G$ starting from $s$.
\end{definition}

\section{ES Trees}\label{sec:estree}
One of the foundational pieces of work was the ES tree by Even and Shiloach~\cite{JACM:EveShi81}, which solves the 
decremental connectivity problem, on unweighted, undirected graphs. The algorithm has constant query time and $\bigO{q+m\cdot n}$ 
total update time, for $q$ queries. Hence for $q > n$, the algorithm outperforms the naive solution of rerunning 
a static algorithm after each edge deletion (which has total runtime $\bigO{q\cdot m}$). This was the first algorithm to beat the naive solution. It has since
had many generalisations~\cite{FOCS:HenKin95,FOCS:King99,SODA:BerRod11}. We begin by discussing the original algorithm before giving the generalisations.

\subsection{Original~\cite{JACM:EveShi81}}
In this section we will discuss the algorithm by Even and Shiloach~\cite{JACM:EveShi81}, before giving an example of how it behaves on a small graph.

Given a graph $G=(V,E)$, the algorithm stores an array which states which connected component each vertex is in. This
array can be used to answer connectivity queries in $\bigO{1}$ time and thus all that is required is to show that
the array can be maintained in $\bigO{m\cdot n}$ total time.

The algorithm will construct a shortest path graph $H$, starting from an arbitrary root vertex $r$. We construct a distance
oracle $d$ such that $d(r)=0$. Any vertex $v$ at distance $i$ from $r$ is assigned $d(v)=i$. This can be calculated using
a Breadth First Search. $u$ is said to be a witness of $v$, if the edge $(u,v)$ is in $H$ and $d(u)=i-1$. If there is a component of
the graph which is not connected to $r$, choose a vertex $r'$, assign it $d(r')=1$ by adding an artificial edge $(r,r')$
and continue with the BFS, repeating until the whole graph is within the structure. This can be achieved in $\bigO{m+n}$ time.

When an edge $(u,v)$ is removed, there are two cases: edges $u$ and $v$ continue to belong to the same connected component, or they
now belong to different connected components. Two processes will be run in parallel\footnote{Here parallel refers to running the two
processes in an interleaved fashion.}, one to deal with each of these cases. Each process is discussed in turn. The processes will `race',
so the first one to finish, will terminate the other one. Thus, we only need to consider the time complexity of the process when it finds
the event (if the graph is still connected or not) that it was looking for.

{\bf Process One} checks whether removing $(u,v)$ disconnects the two components and handles this case. It does so by
calling two Depth First Searches (DFS) from $u$ and
$v$, on the graph $G$. If either DFS finds the other vertex, \textsc{Process One} stops because they are still connected.\footnote{\textsc{Process Two}
is left to finish, since it must maintain its internal data structure.} However, if one of the DFS finish without
finding the other vertex, the two vertices have become disconnected. The smaller component (the one whose DFS finished first) is given a new
component name in the array. Since each time the smaller component is renamed, using a charging argument on the edges, it can be shown that the
total time complexity for this process is $\bigO{m\cdot\log{m}}$.

{\bf Process Two} handles the second case where removing $(u,v)$ does not disconnect the two components and maintains the shortest path graph $H$.
\textsc{Process Two} runs in parallel with \textsc{Process One} and starts with the assumption that we are in case two. As we will discuss, its actions will be
reversed if \textsc{Process One} determines that we are in-fact in case one.

If $(u,v)\in G\setminus H$ then
$(u,v)$ is simply removed from $G$, hence we only need to consider the case where $(u,v)\in H$. If $d(u)=d(v)$ then removing the
edge $(u,v)$ does not change the connected components. Therefore, assume, without loss of generality $d(v)=d(u)+1$. The function can only be
at most one different for $d(u)$ and $d(v)$ since it is defined as the distance to $r$ and there was an edge $(u,v)$. If there is another witness
$w$ for $v$, simply remove $(u,v)$ from both $H$ and $G$. We will now consider the case where $u$ was the final witness for $v$.

When $(u,v)$ is removed from $H$, $v$ must increase $d(v)$ by at least one. As a side effect of removing $v$, anything rooted at $v$ will also need
to be reinserted into $H$. Starting with $i=d(v)$, repeat the following. For each $w\in V$ with $d(w)=i$ and no incoming edges, remove all outgoing
edges from $w$ and increment $d(w)$ by 1. For each $w$ incremented, if there exists a $y$ such that $d(y)=i$ and $(y,w)\in E$, insert the edge into
$H$, thus adding $w$ back into $H$. Then increment $i$ and repeat until all vertices have been added back to the tree.

Clearly if removing $(u,v)$ does not disconnect the graph, then this process will terminate. If $(u,v)$ does disconnect the graph, then \textsc{Process One}
can detect this and cancel \textsc{Process Two} which can reset the data structure to its original state and simply mark the edge $(u,v)$
as artificial. If the graph remains connected the algorithm runs in total time $\bigO{m\cdot n}$ - each time an edge is processed one of its ends
drops by a level. Since $d(v)< n\ \forall v\in V$, each edge can be processed at most $\bigO{n}$ times. This gives the desired time complexity.

Figure~\ref{fig:ESexample} gives an example of the construction, on a given graph. The left hand column shows how the graph $G$ changes over a
sequence of deletions. The middle column shows $H$ as it corresponds to $G$, when it is created with $A$ as the root vertex. The dashed lines
represent edges which are in the original graph but sit in the same level of the $H$. The right hand column shows when only a shortest path
tree $T$ is stored instead of the shortest path graph, this will be discussed in more detail below.

Figure~\ref{fig:ESfalling} shows how the data structures change upon the deletion of the edge $(A,B)$ -- the final edge deletion in
Figure~\ref{fig:ESexample}. In the graph $G$ the edge $(A,B)$ is removed and nothing else changes, hence, the figure focuses on the
steps undergone by $H$ and $T$ (which are identical up until the final step).\footnote{\textsc{Process One} will not succeed and
thus will not be discussed here.} The first step is to remove the edge $(A,B)$ from the data structure.
As $B$, which is in level 1, has no edges connecting it to level 0 ($A$), it is dropped by a level. The data structure $H$ directly has
access to this information, while $T$ has to use the adjacency information of $G$. At the next step (looking at level 2), both $B$ and $D$
need to be considered. The vertex $B$ has no edges to the level above and thus drops a level. In $H$, $D$ is connected to level 1 and
thus is done, while in $T$ it is not connected, so the adjacency information is checked and the edge $(C,D)$ is added to $T$. Finally, $B$
would be checked in level 3 and is connected to level 2 and thus the process completes.

\input{graphic}
\input{graphic-falling}

\subsection{Generalisation}
In this section we discuss how the above can be generalised to other forms of graph or problem.
\subsubsection{Directed graph}
Henzinger and King show that a similar approach can be taken for single source transitivity~\cite{FOCS:HenKin95}. It will only consider the connected component the source $s$ is in
and construct the ES tree using it as the root. The ES tree will now only contain vertices in the same connected component as $s$
and will not contain any artificial edges. Other than this minor modification the algorithm behaves as previously described.

Single sink transitivity problems can be answered in a similar manner, by reversing the direction of all edges before constructing
the ES tree. Transitivity for arbitrary vertex pairs can be answered by storing  an ES tree per vertex.

We use $out(x)$ to denote the ES tree with root $x$ (single source) and $in(x)$ to denote the ES tree with root $x$
on the graph with all the edge directions reversed (single sink).

\subsubsection{Shortest path}
Henzinger and King also show how to answer single source shortest path problems~\cite{FOCS:HenKin95}. An ES tree can be created using the source $s$ as the root of the tree. However,
the ES tree will only be constructed for the connected component $s$ is in, it will not contain any artificial edges or vertices
not in the same connected component. The distance between $s$ and a queried vertex $u$ is then simply $d(u)$ and the path can be
constructed using the ES tree.

Single sink shortest path problems can be answered in a similar manner, by reversing the direction of all edges before constructing
the ES tree. All pairs shortest paths can be answered by storing  an ES tree per vertex.

\subsubsection{Weighted graph}
King shows how the above descriptions can be adjusted to deal with weighted graphs~\cite{FOCS:King99}.
Again here we will focus on a single connected component and can boost
the result by storing multiple ES trees.
The function $d$ now represents the distance of a vertex from the root. The graph will only be stored up to a given depth $\Delta$.
This will result in a total update time complexity of $\bigO{m\cdot\Delta}$. Note, when $\Delta=n$ we get back the original result.

It may no longer be true that $d(v)=d(u)+1$ since $\ell(u,v)$ might be greater than 1. In fact, $d(v)=d(u)+\ell(u,v)$, for $(u,v)\in E$.
Therefore, when a vertex $w$ is moved down from $i$ to layer $i+1$ it won't just be vertices in layer $i$ which are considered
but all vertices $y$ such that $y\in H,(y,w)\in E$ and seeing if $d(y)+\ell(y,w)=i+1$. Note that this does not change the
time complexity. It is straightforward to see the result still holds.

We use $out(x,\Delta)$ to represent the ES tree when the distance is restricted to $\Delta$, $in(x,\Delta)$ is defined similarly.

\subsubsection{Limited Insertions}
While ES trees do not provide a fully dynamic algorithm, Bernstein and Roditty show how the algorithm can handle a very
specific form of insertions~\cite{SODA:BerRod11}. If the insertion of an edge $(u,v)$ does not decrease the distance between the root of the tree and $v$,
then it can be supported by the ES Tree. The total update time is now $\bigO{m'\cdot \Delta + d}$
where $m'$ is the maximum number of edges ever in $G$ and $d$ is the total number of edge changes. The $\bigO{d}$ arises because $\bigO{1}$
work must be spent per edge change. This property will become useful in Sect.~\ref{sec:DDSSSP} to construct a more efficient decremental shortest path algorithm.

This can also be used to increase the weight of a given edge, by inserting the edge, with the new weight, and then deleting the original edge
from the ES tree~\cite{FOCS:King99}.

\subsection{Reduced Memory}
King and Thorup~\cite{ICCC:KinTho01} show how to reduce the space for ES trees, as well as several other algorithms~\cite{FOCS:DemIta00,FOCS:King99}.
The technique allows the memory (beyond the input) to be reduced from $\bigO{m}$ to $\bigO{n}$. Firstly assume an ordering on the vertices.
With such an ordering in place, the algorithm can be tweaked as follows.

Instead of storing all of $H$, the edge $(x,v)$ is stored such that $x$ is first in the vertex ordering of all vertices with the property that $(x,v)\in H$.
Therefore, instead of storing all of $H$, a tree $T$ is stored instead, which has memory requirement $\bigO{n}$. It just remains to show that the tree
can be updated without changing the time complexity.

An edge deletion $(u,v)$ will only matter if it was in $T$, if not it is simply removed from $G$. If $(u,v)\in T$ a new edge must be found,
the edges leading to $v$ can be scanned, in order, starting at $u$, stopping if $(x,v)$ is found such that $d(x)+\ell(x,v)=d(v)$ and adding $(x,v)$ to $T$.
If not the value of $d(v)$ is incremented. As before, each edge into $v$ is only considered once for each value of $d(v)$. Hence, the running
time remains the same.

Several other works have also given space saving techniques~\cite{FOCS:DemIta00,MISC:BroKin00} but these only allow the distance of the shortest path to
be given and can not produce the path. Thus they will not be discussed in detail here.



\section{Algorithms Built upon ES Trees}\label{sec:usinges}
In this section we describe some of the algorithms which use ES trees as a building block to solve the discussed dynamic algorithms.
While there are a host more, such as~\cite{SODA:BerChe16,SODA:HenKriNan14,FOCS:Bernstein09}, we discuss a subset which made significant progress or contain interesting ideas.

\subsection{Approximate Decremental Single Source Shortest Paths~\cite{STOC:BerChe16}}\label{sec:DDSSSP}
In this section we discuss an approximate algorithm for the decremental single source shortest path problem, for unweighted and undirected graphs, by Bernstein and
Chechik~\cite{STOC:BerChe16}. This is the first deterministic algorithm which manages to have a total update time better than
the $\bigO{m\cdot n}$ of ES trees. The algorithm is a $(1+\epsilon)$ approximation with a total update time of $\tO{n^2}$.

Here we will formally describe an algorithm that gives a time complexity of $\bigO{n^{2.5}}$ and then discuss how to improve
this to the stated bound.

A vertex $v\in V$ is called heavy if it has degree at least $\sqrt{n}$ and light otherwise. A path between two vertices can contain at most $5\sqrt{n}$ vertices - intuitively, no two heavy vertices can share a common neighbour else there would be a shorter path. Since a heavy vertex has $\sqrt{n}$ neighbours the result follows. This will be discussed formally below.

The algorithm works by storing two ES trees one which will store short paths and one which will approximate the long paths. For short paths we will simply create a ES tree on the original graph from source $s$ up to depth $5\sqrt{n}\epsilon^{-1}$. The remainder of this explanation will discuss long paths.

Let $H$ be the set of heavy vertices in $G$. The auxiliary graph $G'$ is the graph with an additional vertex $c$ per connected component in $G[H]$ which is connected to each vertex in the connected component with weight $1/2$. The light vertices then have all their edges added to the graph. The graph $G'$ has at most $n^{1.5}$ edges; 1 per heavy vertex and at most $\sqrt{n}$ per light vertex. We will now show that this graph provides the following bounds:

\begin{eqnarray*}
 \mathbf{dist}_{G'}(s,v)\leq\mathbf{dist}_{G}(s,v) \leq \mathbf{dist}_{G'}(s,v) + 5\sqrt{n}
\end{eqnarray*}

The lowerbound follows from the observation that given an $s$-$t$ path in $G$, an $s$-$t$ path 
can be constructed in $G'$ as follows. For each $(u,v)$ if either $u$ or $v$ are light then $(u,v)\in E'$ and can be added to the path. If $(u,v)$ are both heavy then they must be in the same connected component so $(u,c),(c,v)$ can be added to the path. Since these two edges have weight half the pair have the same weight as the original path. Thus the path in $G'$ has the same weight as the path in $G$. Note that this path may not be simple. 

For the upper bound let $L$ be the set of light vertices on the shortest path in $G'$ and let $X$ be the set of non light edges on the path (so either heavy or a center $c$). We want to show that $|X|<5\sqrt{n}$, therefore showing that ignoring the heavy edges doesn't cost too much. Let $Y$ be every 5\textsuperscript{th} element of $X$, therefore $|Y|\geq\frac{|X|}{5}$. We know that $Ball(X,v,2)\geq \sqrt{n}$ for $v\in Y$, since either $v$ is heavy or adjacent to a heavy vertex. For $v,w\in Y$ we know $Ball(X,v,2)\cap Ball(X,w,2) = \emptyset$ otherwise there would be a shorter path (since $v$ and $w$ are at least distance 5 away from each other on the path). Thus we know $|\cup_{v\in Y}Ball(X,v,2)|\geq \sqrt{n}|Y|$ but since the graph contains at most $n$ vertices we get the desired result.

An ES tree can be stored for the original graph $G$ up to distance $5\sqrt{n}\epsilon^{-1}$ to respond to short edge queries, while the ES tree up to distance $n$ on $G'$ allows us to respond to longer path queries. Both run in time $\bigO{n^{2.5}}$, the first due to its bounded depth and the second because it is sparse.

We now need to show that the distances can be maintained under edge deletions. Any edge incident to a light vertex is easy to maintain as it is in the original graph. Deleting an edge can cause a vertex to go from heavy to light (but this can only happens once). When this happens, all of its edges must be added to the auxiliary graph. The slightly trickier case is when an edge is deleted between two heavy vertices. A data structure which maintains connectivity information in dynamic graphs can be used to maintain the auxiliary graph. When edge $(u,v)$ is deleted it must be checked that $(u,v)$ are still in the same connected component. If yes nothing changes. Else these two edges now need to be connected to different centers $c_u,c_v$ instead of the same center $c$. This is done by choosing the smallest center and moving all vertices adjacent to it over to a newly created center. Hence, the graph can be maintained. 

To reduce the time complexity from $\tO{n^{2.5}}$ to $\tO{n^2}$, instead of having two ES trees (a ``heavy one'' and a ``light one''), $\bigO{\log{n}}$ heaviness thresholds can be used to handle $\bigO{\log{n}}$ ranges of distance queries and returning the minimum of the $\bigO{\log{n}}$ queries.

The algorithm can be trivially converted to the incremental setting, with the same time complexity.

\subsection{Fully Dynamic Transitivity~\cite{FOCS:HenKin95}}

Henzinger and King~\cite{FOCS:HenKin95} give the first fully dynamic transitive closure algorithm, along with a decremental algorithm.
Both are Monte Carlo algorithms. The fully dynamic algorithm has either; query time $\bigO{\frac{n}{\log{n}}}$ and update time
$\bigO{\hat{m}\cdot\sqrt{n}\cdot\log^2{n}+n}$, or query time  $\bigO{\frac{n}{\log{n}}}$ and update time $\bigO{n\cdot\hat{m}^{\frac{\mu-1}{\mu}}\cdot\log^2{n}}$
where $\hat{m}$ is the average number of edges in the graph and $\mu$ is the exponent for matrix multiplication. Note that, unlike
the other algorithms given, these algorithms do not have a constant query time.

The deletions only algorithm takes in a user defined parameter $r$ and for $i=1,\ldots,\log{r}$ stores a set of $min(2^i\cdot\log{n},n)$
distinguished vertices $S_i$. For each distinguished vertex $x$ maintain ES trees $in(x,\frac{n}{2^i})$ and $out(x,\frac{n}{2^i})$, where $x\in S_i$.
Then $out(x)$ can be defined as the union of all $out(x,\frac{n}{2^i})$ where $x\in S_i$, with $in(x)$ being defined similarly. For each
vertex $v\in V$ also maintain $in(v,\frac{n}{r})$ and $out(v,\frac{n}{r})$

Given a query $(u,v)$, test if $v$ is in $out(u,\frac{n}{r})$, if yes return true. Else see if there exists a distinguished vertex $x$ such
that $u\in in(x)$ and $v\in out(x)$. If this is the case answer yes, else answer no.

If the path is of length less than $\frac{n}{r}$ then the answer will always be correct, otherwise it will be correct with high probability.

The two fully dynamic algorithms use similar techniques of using the deletion only data structure discussed above and suitably
keeping track of inserted edges. Thus only one of the two will be discussed here. The intuition is that you store the deletion only
data structure, for $r=\frac{n}{\log^2{n}}$, along with storing $in(x,n)$ and $out(x,n)$ each time an edge is inserted. These
extra structures, along with the deletion only data structure are updated each time an edge is removed. To answer a query $(u,v)$,
query the deletion only data structure and query if $u\in in(x,n)$ and $v\in out(x,n)$ for every newly inserted edge. After every $\sqrt{n}$
updates to the graph the deletion only data structure is rebuilt. This gives the desired result.

\subsection{Fully Dynamic All Pairs Shortest Paths and Transitivity~\cite{FOCS:King99}}
King~\cite{FOCS:King99} gives the first fully dynamic algorithms for all pairs shortest paths in directed and weighted graphs, where the weight is bounded by some positive integer $b$.
Three algorithms are given; a $(2+\epsilon)$ approximation, a $(1+\epsilon)$ approximation and an exact algorithm with amortized update times;
$\bigO{\frac{n^2\cdot\log^2{n}}{\log{\log{n}}}},\bigO{\frac{n^2\cdot\log^3{(b\cdot n)}}{\epsilon^2}}$ and $\bigO{n^{2.5}\sqrt{b\cdot\log{n}}}$
respectively. They also give a fully dynamic transitive closure algorithm with update time $\bigO{n^2\cdot\log{n}}$. The update times are
amortized over $\bigO{\frac{m}{n}}$ operations.

Here we describe the transitive closure algorithm, since the others follow a similar strategy. The algorithm works by keeping $k=\lceil\ln{n}\rceil$
forests, denoted $F^1,\ldots,F^k$, such that $F_i$ contains $in^i(v,2)$ and $out^i(v,2)$ for each $v\in V$. The count between two vertices
$count^i(u,w)$ is the number of vertices $v$ such that $u\in in^i(v,2)$ and $w\in out^i(v,2)$. The list $list^i(u,w)$ contains all such vertices.
The forests (and the graphs they are built upon) are defined recursively. The edges $E^i=\{(u,w)|count^{i-1}(u,v)>0\}$ and then the forests
are the in and out ES trees on top of this. Inserting an edge requires adding it at the bottom layer and adjusting the edge sets and forests
as required. Deleting the edge requires deleting it from the lowest layer and the recursively deleting edges in higher layers where the count
has gone from positive to zero (\ie where there is no longer a path). It follows that $u,w\in V$ are connected in $G$ if and only if
$count^k(u,w)>0$, resulting in a transitive closure algorithm.

\subsection{Decremental Single Source Shortest Paths~\cite{FOCS:HenKriNan14}}
Henzinger \etal~\cite{FOCS:HenKriNan14} present an algorithm for $(1+\epsilon)$ decremental single-source shortest path for
unweighted graphs, with total update time $\bigO{m^{1+o(1)}}$. The algorithm works by maintaining a sparse $(d,\epsilon)$-hop set
(introduced by Cohen~\cite{JACM:Cohen00}). This allows the distance between any two vertices to be $(1+\epsilon)$-approximated using
at most $d$ edges. To maintain the hop set, under deletions, the authors introduce a \emph{monotone bounded-hop Even Shiloach tree}.

The high-level idea of the algorithm is to create a hop set and then defining the shortcut graph as the original graph
plus the edges from the hop set. This process is the repeated on the resulting graph up to a suitable depth. Intuitively
this works because while the graph gains more edges, the number of hops is being constrained. The final algorithm
is slightly more complex, where it has to contain active and inactive vertices. Inactive vertices are those which it would be too expensive
to maintain an ES-tree for but the authors show that not constructing these trees will not change the result. See the paper for the
formal definition of active and inactive vertices.

\subsection{Decremental Single Source Shortest Paths~\cite{STOC:HenKriNan14}}
Henzinger \etal~\cite{STOC:HenKriNan14} give a $(1+\epsilon)$-approximation decremental single-source shortest path algorithm for directed graphs with total update
time $\tO{m\cdot n^{0.984}}$ and constant update time. Here we will describe the $s$-$t$ reachability algorithm, and the authors show
how it can be extended to single source reachability and shortest paths.

Given a set of vertices $H$ called a hub, the hub distance between $u$ and $v$ is the shortest distance between $u$ and $v$ such
that one of the vertices in the path belongs to $H$. A path between $u$ and $v$ is called a $h$-hop if it contains at most $h$
edges. The $h$-hop $u$-$v$ path union graph is the graph created by taking the union of all the $h$-hops between $u$ and $v$. The goal
of the algorithm is to maintain reachability while $dist(s,t)\leq h$ for some parameter $h$. The algorithm then maintains the hub distance between $s$ and $t$
while the hub distance is less than $h$ and maintains the distance between $s$ and $t$ in the path union graph when the distance becomes
greater than $h$.

For each $v\in H$ $dist(s,v)$ and $dist(v,t)$ are maintained using ES-trees of depth $h$. Maintaining the distance between $s$ and $t$ in
the path union graph varies depending on the algorithm given in the paper. The authors give algorithms for sparse graphs, dense graphs and
other graphs. One way to do this is maintaining an ES tree over the path union graph.

The authors later improved upon this~\cite{ArXiv:HenKriNan16} for the single source reachability case, with total update time
$\bigO{m\cdot n^{0.98+o(1)}}$. The algorithm works by extending to multiple
layers of path union graphs and hubs from the previous algorithm.

\section{Algorithms Using Alternate Techniques}\label{sec:noes}
In this section we discuss some other algorithms which do not utilise ES trees. While there is a host of work, such as~\cite{ARXIV:PonRam14,AAM:DinLin14,ISAC:NasPonRam14},
we focus on a particular subset which contain interesting concepts.

\subsection{Approximate Decremental All Pairs Shortest Paths~\cite{SODA:BerRod11}}

Bernstein and Roditty~\cite{SODA:BerRod11} give the first decremental algorithms to beat the $\bigO{m\cdot n}$ update time of ES trees.
These algorithms beat the time complexity only when the graph is ``not too sparse''. They present two randomised algorithms for shortest
path problems on unweighted and undirected graphs.

The first algorithm is a $(1+\epsilon)$ approximation algorithm for single source shortest path, with an expected total update
time of $\tO{n^{2+\bigO{\frac{1}{\sqrt{\log{n}}}}}}$. The second algorithm just returns the distances for the all pairs shortest path problem with total
expected update time of $\tO{n^{2+\frac{1}{k}+\bigO{\frac{1}{\sqrt{\log{n}}}}}}$ for any fixed integer $k$, where
the returned distances are at most a $2\cdot k -1 + \epsilon$ approximation of the true distance.

Both algorithms are created using a similar technique. Existing algorithms are run on a sparse subgraph of the graph (\eg spanner or emulator) which
is being queried, instead of the graph itself. This technique has been used previously to construct  more efficient static algorithms~\cite{JOC:ABCP98,JOC:Cohen98,JOC:ACIM99,JOC:DorHalZwi00,JOA:CohZwi01,JOC:ElkPel04,PODC:Elkin01,JACM:Zwick02,JACM:ThoZwi05}.
However, it is more complex in the decremental only setting. The issue arises, in that a deletion
from the original graph, can cause an insertion into the emulator. The authors resolve the issue by showing that the insertions
will be limited and ``suitably well behaved''. Given this they are able to show that existing decremental algorithms can support
the required insertions, giving the desired result.

\subsection{Dynamic All Pairs Shortest Paths~\cite{STOC:DemIta03}}
Demetrescu and Italiano~\cite{STOC:DemIta03} give a, deterministic, fully dynamic algorithm for APSP on directed graphs with non-negative real-valued edge weights. The algorithm
has amortized update time $\bigO{n^2\cdot\log^3{n}}$ and worst case constant time query time. We begin by discussing the algorithm
which can only increase the weight edges and has amortized update time $\bigO{n^2\cdot\log{n}}$. This was the first algorithm to do better than simple recomputation.

An important definition for this problem is that of a local shortest path. A local shortest path is a path where all proper subpaths
are shortest paths. Note this does not require that the path itself is a shortest path. The algorithm works by storing priority queues
$P_{x,y}$ of locally shortest paths between $x$ and $y$, and $P^*_{x,y}$ of shortest paths between $x$ and $y$. Distance queries are
answered by returning elements from the correct priority queue. It remains to show that the priority queues can be updated efficiently.

The paper shows that a graph can have at most $m\cdot n$ locally shortest paths (assuming unique shortest paths) and that at most
$\bigO{n^2}$ paths can stop being locally shortest per edge weight increase. Using this result the data structures can be maintained
in the claimed time.

To upgrade the algorithm to be fully dynamic in the claimed time, the authors use historically shortest paths. At current time $t$ such
that the given path had last been updated at time $t'\leq t$, a path is called historical if it has been a shortest path in the interval $[t', t]$.
A path is locally historical if it contains a single vertex or all proper subpaths are historical. The algorithm has an additional step
to keep the number of historical paths at a reasonable level. If there are either too few or too many, it slows the algorithm. Otherwise
the algorithms behaves in a similar manner.

\subsection{Decremental Strongly Connected Components~\cite{TALG:Lac13}}
{\L}\k{a}cki~\cite{TALG:Lac13} gives a deterministic algorithm for strongly connected components, with total update time $\bigO{m\cdot n}$.
The algorithm works by reducing the problem to solving connectivity in a set of directed acyclic graphs. The algorithm begins by removing
all vertices that are not reachable from the given source $s$, which takes $\bigO{m}$ time. When an edge $(u,v)\in E$ is deleted, the
graph becomes disconnected if $v$ loses its last in edge. Not only may this cause $v$ to become disconnected, it may also cause children of
$v$ to become disconnected. The algorithm works using this observation. It starts by taking the vertex $v$ and if it is has no other incoming
edges it declares that $v$ has become disconnected. If $v$ has become disconnected, it must recurse on the all children $x$ of $v$, removing the edge
$(v,x)$ since that no longer connects $x$ to the source. The runtime is linear in the number of newly disconnected vertices and their incident edges.
A similar algorithm can be given for the reachability of a sink by reversing all the edges.

The strongly connected components algorithm works by splitting a vertex $d$ into two $d_1$ and
$d_2$ where all edges $(u,d)\in E$ get replaced with $(u,d_1)$ and edges $(d,v)\in E$ get replaced with $(d_2, v)$. Then the condensation
of the resulting graph is calculated. This graph is denoted $G_d$.
This graph is a DAG and therefore reachability from $d_2$ and to $d_1$ can be maintained using the above algorithm. As soon as one of the reachability
sets is different from the vertex set then the graph has stopped being strongly connected. To handle deletion of edges not in $G_d$ this must be applied
recursively to each strongly connected component, this is stored as a tree.

Roddity~\cite{SODA:Roditty13} improves on the preprocessing and worst case update time of the above algorithm to $\bigO{m\cdot\log{n}}$,
without changing the query time or total update time. This is achieved by giving a new preprocessing algorithm which generates the
tree containing the strongly connected components in $\bigO{m\cdot\log{n}}$. The update function is the original one by {\L}\k{a}cki,
with the adjustment that if it takes over $\bigO{m\cdot\log{n}}$ time then the process is terminated and the tree is simply built again
from scratch.

\section{Limitations}\label{sec:limits}
In this section we discuss work that attempts to give lowerbounds on the complexity of these algorithms~\cite{STOC:Patrascu10,FOCS:AbbWill14,SODA:KopPerPor16,STOC:HKNS15}.
They achieve this by giving
a reduction to a well studied problem. If these dynamic problems can be solved ``too efficiently'' then it would result in a more efficient
algorithm for a problem which is well studied and conjectured that no algorithm can beat a certain threshold. Here we detail the result of
Henzinger \etal~\cite{STOC:HKNS15} who use online boolean matrix-vector multiplication as their underlying problem. This work is for undirected
and unweighted graphs. We chose to discuss the work of Henzinger \etal~because lots of the results given subsume previously known results~\cite{STOC:Patrascu10,FOCS:AbbWill14,SODA:KopPerPor16}.

We begin by defining the online boolean matrix-vector multiplication problem that they reduce all the problems to, to construct lower bounds.

\begin{definition}[Online Boolean Matrix-Vector Multiplication Problem (OMv)~\cite{STOC:HKNS15}]
 Let $n$ be an integer. Given an $n\times n $ Boolean matrix $M$, and, for each of $n$ rounds, an $n$-dimensional column vector $v_i$,
 compute $Mv_i$. The algorithm must output the result before being given the next column vector.
\end{definition}

The OMv problem was chosen because it has been well studied~\cite{AMS:Motzkin55,SODA:Willians07,JOC:Savage74,C:SanUrr86,IPS:LibZuc09}.
It is widely believed that (up to logarithmic
factors) the OMv problem can not be solved more efficiently than $\bigO{n^3}$. This led to the formalisation of the OMv
conjecture which is given below.

\begin{definition}[Online Boolean Matrix-Vector Multiplication Conjecture~\cite{STOC:HKNS15}]
 For any constant $\epsilon>0$, there is no $\bigO{n^{3-\epsilon}}$ time algorithm that solves OMv with an error probability of at most $\frac{1}{3}$.
\end{definition}

For the remainder of the paper, the authors construct lowerbounds for a host of problems by assuming the OMv conjecture to be true~\cite{STOC:HKNS15}. We will describe
the lowerbounds for the problems we are interested in below. However, many more lowerbounds are given, including bounds for;
triangle detection, $d$-failure connectivity, Langerman's problem, Erickson's problem, approximate diameter and densest subgraph.

For what follows, let the preprocessing time, update time and query time be denoted as $p(n), u(n), q(n)$ respectively. We are now in a position to detail the lowerbound results of Henzinger \etal~\cite{STOC:HKNS15}.

\begin{theorem}[\cite{STOC:HKNS15}, Corollary 4.2]
 Assuming the OMv conjecture; for any $n$ and $m=\bigO{n^2}$, there is no decremental $s$-$t$ shortest path with polynomial preprocessing
 time, $u(m)=\tto{m^{\frac{3}{2}}}$ (total) and $q(m)=\tto{m}$.
\end{theorem}

This theorem, in particular, shows why it was many years for an improvement to be given to the algorithm of Even and Shiloach~\cite{JACM:EveShi81}. If $m\in\bigO{n^2}$
then the two bounds are matching. This shows why it is important to consider approximate algorithms instead of exact algorithms. Since the result is for exact algorithms,
considering approximate algorithms could be a way to skirt around this. However, Henziner \etal~also provide some bounds for approximate algorithms.

\paragraph{Note:}  The above theorem also holds for incremental algorithms.

\begin{theorem}[\cite{STOC:HKNS15}, Corollary 3.10]
 Assuming the OMv conjecture;
 there is no decremental algorithm for $(3-\epsilon)$-$s$-$t$ shortest path with 
 $p(n)=poly(n)$, $u(n)=\tto{\sqrt{n}}$ (in the worst case) and $q(m)=\tto{n}$.
\end{theorem}

\paragraph{Note:}  The theorem also holds for incremental algorithms, and fully dynamic algorithms with amortized update time.

We conclude this section by giving a few more results which are also of relevance.

\begin{theorem}[\cite{STOC:HKNS15}, Corollary 3.4]
 Assuming the OMv conjecture; for any $n$ and $m\leq n^2$, there is no partially dynamic algorithm for Strongly Connected Components or Transitivity with
 time $p(m)=poly(m)$, $u(m)=\tto{\sqrt{m}}$ (worse case) and $q(m)=\tto{m}$.
\end{theorem}

\begin{theorem}[\cite{STOC:HKNS15}, Corollary 4.8]
 Assuming the OMv conjecture, for any $n$ and $m=\Theta(n^{\frac{1}{1-\delta}})$, and constant $\delta\in(0,\frac{1}{2}]$, 
 there is no partially dynamic algorithm for the problems listed below with;
 $p(m)=poly(m)$, $u(m)=\tto{m\cdot n}$ (total) and $q(m)=\tto{m^{\delta}}$. The problems are:
 \begin{itemize}
  \item Single Source Shortest Path
  \item All Pairs Shortest Path (2 vs 4)
  \item Transitive closure
 \end{itemize}
\end{theorem}

The 2 vs 4 variant of the APSP problem requires you to distinguish if the shortest path between the given vertices is less than, or equal to, two
or if it is greater than, or equal to, four. This shows that even decision variants of these problems can't be solved ``too efficiently''.

Recently Larsen and Williams~\cite{ARXIV:LarWil16} showed that a proof of the OMv conjecture could not be constructed only using information
theoretic arguments. However, it is still possible for the conjecture to hold.

\section{Conclusion and Open Problems}\label{sec:open}

In this paper we survey, both the foundational and state of the art, algorithms for computing shortest paths and connectivity
on graphs which are constantly changing. Table~\ref{tbl:upper} summarises the time complexities of all the algorithms discussed
in this paper, while Table~\ref{tbl:lower} gives the known lower bounds. We now provide a brief conclusion and describe some of the open problems.\footnote{All
individual work posed their own open questions but here we try to focus on themes more than individual questions.}


\paragraph{Derandomisation} Currently the randomised algorithms tend to perform better than the deterministic algorithms.
However, as discussed in the introduction randomised algorithms require a weaker adversary, who does not know the randomness and can not make
adaptive queries. For real world scenarios, it is important to be able to remove these restrictions. This makes deterministic algorithms
more desirable, since they do not have these restrictions. Hence, an important open question is if the randomised algorithms can be
derandomised or if deterministic algorithms can be constructed with the same time complexities as their randomised counterparts.

\paragraph{Memory} To have a constant query time for all pairs shortest path and transitive closure, at least $\Omega(n^2)$ memory is
required for the lookup table. For connectivity and single source/sink shortest path only $\Omega(n)$ memory is required. However, lots of
the dynamic algorithms require more space than this. Thus an important question is if existing algorithms can have their memory reduced to
the lowerbound or if new algorithms can be designed meeting this bound. As mentioned above, there has been work moving algorithms in this
direction~\cite{ICCC:KinTho01,FOCS:DemIta00,MISC:BroKin00}

\paragraph{Lower Bounds}It is an important question to try and prove the online matrix vector multiplication conjecture, or equivalently conjectures
from other lowerbound work. As discussed by Henzinger \etal~\cite{STOC:HKNS15}, the lower bound techniques tend to work for both incremental and decremental algorithms. Therefore, it is an interesting question to
see if these can be bounded separately, to get tighter bounds. For example, incremental single source connectivity can be solved in time $\bigO{m\cdot\alpha(n)}$
while the best known result for the decremental setting is $\bigO{m\cdot n^{0.98+o(1)}}$. In relation to the derandomisation question
above, to show a seperation between deterministic and randomised algorithms, would require applying new techniques which only hold for the deterministic setting.

\subsubsection{Acknowledgements}
The author is extremely grateful to Benjamin Sach for proof reading and helpful discussions.

\begin{landscape}
\begin{table}[t]
 \centering
 \begin{tabular}[t]{|c|c|Hc|c|Hc|c|HHHc|c|}\hline
   Algorithm&Dynamic&$p(m,n)$&$u(m,n)$&$q(m,n)$&Memory&Weighted&Directed&$p(m,n)$&$u(m,n)$&$q(m,n)$&Time Complexity & Approximation\\\hline
   \multicolumn{13}{|c|}{Transitive Closure}\\\hline
   \cite{JACM:EveShi81}&Dec&$m$&$m\cdot n^2$&1&$m\cdot n$&\xmark&\xmark&\multirow{4}{*}{poly}&\multirow{4}{*}{$m\cdot n^{1-\epsilon}$}&\multirow{4}{*}{$m^{\delta'-\epsilon}$}&Total&\multirow{4}{*}{N/A}\\\cline{1-8}\cline{9-12}
   \cite{FOCS:HenKin95}&Fully&$?$&$\hat{m}\sqrt{n}\cdot\log^2{n}+n$&$\frac{n}{\log{n}}$&Mem&\xmark&\xmark&&&&Amortized&\\\cline{1-8}\cline{9-12}
   \cite{FOCS:HenKin95}&Fully&$?$&$n\cdot\hat{m}^{\frac{\mu-1}{\mu}}\cdot\log^2{n}$&$\frac{n}{\log{n}}$&Mem&\xmark&\xmark&&&&Amortized&\\\cline{1-8}\cline{9-12}
   \cite{FOCS:King99}&Fully&$?$&$n^2\cdot\log{n}$&1&Mem&$\mathbb{Z}^+_{<b}$&\cmark&&&&Amortized&\\\hline
   \multicolumn{13}{|c|}{Strongly Connected Components}\\\hline
   \cite{TALG:Lac13}&Dec&$?$&$m\cdot n$&1&Mem&\cmark&\cmark&\multirow{2}{*}{poly}&\multirow{2}{*}{$m\cdot n^{1-\epsilon}$}&\multirow{2}{*}{$m^{\delta'-\epsilon}$}&Total&\multirow{2}{*}{N/A}\\\cline{1-8}\cline{9-12}
   \cite{SODA:Roditty13}&Dec&$?$&$m\cdot\log{n}$&1&Mem&\cmark&\cmark&&&&Worst case (Total as above)&\\\hline
   \multicolumn{13}{|c|}{Single Source Shortest Path}\\\hline
   \cite{JACM:EveShi81}&Dec&$m$&$m\cdot n$&1&$m$&\xmark&\xmark&\multirow{6}{*}{poly}&\multirow{6}{*}{$m\cdot n^{1-\epsilon}$}&\multirow{6}{*}{$m^{\delta'-\epsilon}$}&Total&1\\\cline{1-8}\cline{9-13}
   \cite{STOC:BerChe16}&Dec&$m$&$n^2$&1&Mem&\xmark&\xmark&&&&Total&$1+\epsilon$\\\cline{1-8}\cline{9-13}
   \cite{FOCS:HenKriNan14}&Dec&$m^{1+o(1)}$&$m^{1+o(1)}$&1&Mem&$\log{W}$ slow down&\xmark&&&&Total&$1+\epsilon$\\\cline{1-8}\cline{9-13}
   \cite{STOC:HenKriNan14}&Dec&$?$&$m\cdot n^{0.984}$&1&Mem&$\mathbb{Z}^+_{2^{\log^c{n}}}$&\cmark&&&&Total&$1+\epsilon$\\\cline{1-8}\cline{9-13}
   \cite{ArXiv:HenKriNan16}&Dec&$?$&$m\cdot n^{0.98+o(1)}$&1&Mem&$\mathbb{Z}^+_{2^{\log^c{n}}}$&\cmark&&&&Total&$1+\epsilon$\\\cline{1-8}\cline{9-13}
   \cite{SODA:BerRod11}&Dec&$?$&$n^{2+\bigO{\frac{1}{\sqrt{\log{n}}}}}$&1&Mem&\xmark&\xmark&&&&Total&$1+\epsilon$\\\hline
   \multicolumn{13}{|c|}{All Pairs Shortest Path}\\\hline
   \cite{JACM:EveShi81}&Dec&$m$&$m\cdot n^2$&1&$m\cdot n$&\xmark&\xmark&\multirow{6}{*}{poly}&\multirow{6}{*}{$m\cdot n^{1-\epsilon}$}&\multirow{6}{*}{$m^{\delta'-\epsilon}$}&Total&1\\\cline{1-8}\cline{9-13}
   \cite{FOCS:King99}&Fully&$?$&$\frac{n^2\cdot\log^2{n}}{\log{\log{n}}}$&1&Mem&$\mathbb{Z}_{<b}^+$&\cmark&&&&Amortized over $\frac{m}{n}$&$2+\epsilon$\\\cline{1-8}\cline{9-13}
   \cite{FOCS:King99}&Fully&$?$&$\frac{n^2\cdot\log^3{(b\cdot n)}}{\epsilon^2}$&1&$n^3$&$\mathbb{Z}_{<b}^+$&\cmark&&&&Amortized over $\frac{m}{n}$&$1+\epsilon$\\\cline{1-8}\cline{9-13}
   \cite{FOCS:King99}&Fully&$?$&$n^{2.5}\sqrt{b\cdot\log{n}}$&1&$n^3$&$\mathbb{Z}_{<b}^+$&\cmark&&&&Amortized over $\frac{m}{n}$&$1$\\\cline{1-8}\cline{9-13}
   \cite{SODA:BerRod11}&Dec&$?$&$n^{2+\frac{1}{k}+\bigO{\frac{1}{\sqrt{\log{n}}}}}$&1&$m+n^{1+\frac{1}{k}}$&\xmark&\xmark&&&&Total&$2\cdot k - 1 +\epsilon$\\\cline{1-8}\cline{9-13}
   \cite{STOC:DemIta03}&Fully&$?$&$n^2\cdot\log^3{n}$&1&Mem&$\mathbb{R}^+$&\cmark&&&&Amortized&1\\\hline
   \cite{FOCS:RodZwi04}&Fully&$?$&$\frac{m\cdot n}{t}$&$t$&Mem&\xmark&\xmark&&&&Amortized&$1+\epsilon$\\\hline
   \cite{FOCS:RodZwi04}&Dec&$?$&$m\cdot n$&$1$&Mem&\xmark&\xmark&&&&Expected&$1+\epsilon$\\\hline
   \cite{SODA:BasHarSen03}&Dec&&$n^{\frac{10}{9}}$&1&1&\xmark&\xmark&&&&Amortized&$3$\\\hline
   \cite{SODA:BasHarSen03}&Dec&&$n^{\frac{14}{13}}$&1&1&\xmark&\xmark&&&&Amortized&$5$\\\hline
   \cite{SODA:BasHarSen03}&Dec&&$n^{\frac{28}{27}}$&1&1&\xmark&\xmark&&&&Amortized&$7$\\\hline
\end{tabular}
\caption{A summary of known upper bounds for various problems.}
\label{tbl:upper}
\end{table}

\begin{table}[t]
 \centering
 \begin{tabular}[t]{|c|c|c|c|c|c|c|}\hline
 Problem&$p(m,n)$&$u(m,n)$&$q(m,n)$&Assumption&Remark& Citation\\\hline
 \multirow{2}{*}{Transitive Closure}&$(m\cdot n)^{1-\epsilon}$&$(m\cdot n)^{1-\epsilon}$&$m^{\delta-\epsilon}$&BMM&\multirow{2}{*}{$\delta\in(0,\frac{1}{2}],m=\Theta(n^{\frac{1}{1-\delta}})$}&\cite{JOC:DorHalZwi00}\\\cline{2-5}\cline{7-7}
				    &{poly}&$(m\cdot n)^{1-\epsilon}$&$m^{\delta-\epsilon}$&OMv&&\cite{STOC:HKNS15}\\\hline
 Strongly Connected Components&{poly}&{$m\cdot n^{1-\epsilon}$}&{$m^{\delta-\epsilon}$}&OMv&$\delta\in(0,\frac{1}{2}],m=\Theta(n^{\frac{1}{1-\delta}})$&\cite{STOC:HKNS15}\\\hline
 \multirow{3}{*}{Single Source Shortest Path}&poly&{$m^{\frac{3}{2}-\epsilon}$}&{$m^{1-\epsilon}$}&OMv&$m=\bigO{n^2}$&\cite{STOC:HKNS15}\\\cline{2-7}
					     &$(m\cdot n)^{1-\epsilon}$&$(m\cdot n)^{1-\epsilon}$&$m^{\delta-\epsilon}$&BMM&\multirow{2}{*}{$\delta\in(0,\frac{1}{2}],m=\Theta(n^{\frac{1}{1-\delta}})$}&\cite{A:RodZwi11}\\\cline{2-5}\cline{7-7}
					     &{poly}&$(m\cdot n)^{1-\epsilon}$&$m^{\delta-\epsilon}$&OMv&&\cite{STOC:HKNS15}\\\hline
 \multirow{2}{*}{All Pairs Shortest Path}&$(m\cdot n)^{1-\epsilon}$&$(m\cdot n)^{1-\epsilon}$&$m^{\delta-\epsilon}$&BMM&\multirow{2}{*}{$\delta\in(0,\frac{1}{2}],m=\Theta(n^{\frac{1}{1-\delta}})$}&\cite{JOC:DorHalZwi00}\\\cline{2-5}\cline{7-7}
				    &{poly}&$(m\cdot n)^{1-\epsilon}$&$m^{\delta-\epsilon}$&OMv&&\cite{STOC:HKNS15}\\\hline
 \end{tabular}
\caption{A summary of known lower bounds for various problems.}
\label{tbl:lower}
\end{table}
\end{landscape}

\end{document}

%% file: spg.tex
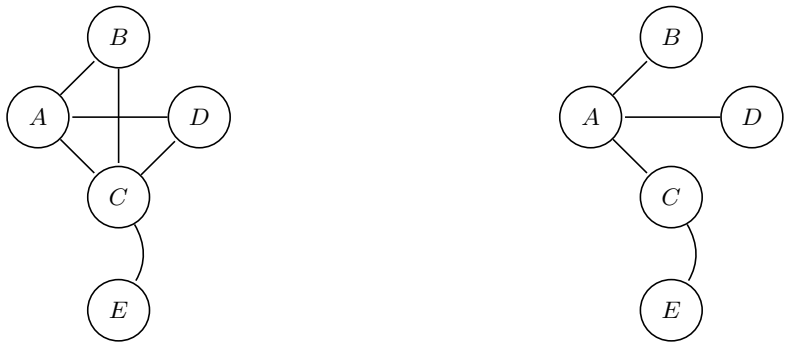
\begin{figure}
\begin{minipage}{0.45\textwidth}
\centering
\begin{tikzpicture}[-,>=stealth',shorten >=1pt,auto,node distance=1.5cm,
                    semithick]
  \tikzstyle{every state}=[fill=none,draw=black,text=black]

  \node[state]         (A)                    {$A$};
  \node[state]         (B) [above right of=A] {$B$};
  \node[state]         (C) [below right of=A] {$C$};
  \node[state]         (D) [below right of=B] {$D$};
  \node[state]         (E) [below of=C]       {$E$};

  \path (A) edge              node {} (B)
            edge              node {} (C)
        (B) edge              node {} (C)
        (C) edge              node {} (D)
            edge [bend left]  node {} (E)
        (D) edge              node {} (A);
\end{tikzpicture}
 \end{minipage}
\begin{minipage}{0.45\textwidth}
\centering
\begin{tikzpicture}[-,>=stealth',shorten >=1pt,auto,node distance=1.5cm,
                    semithick]
  \tikzstyle{every state}=[fill=none,draw=black,text=black]

  \node[state]         (A)                    {$A$};
  \node[state]         (B) [above right of=A] {$B$};
  \node[state]         (C) [below right of=A] {$C$};
  \node[state]         (D) [below right of=B] {$D$};
  \node[state]         (E) [below of=C]       {$E$};

  \path (A) edge              node {} (B)
            edge              node {} (C)
        (C) edge [bend left]  node {} (E)
        (D) edge              node {} (A);
\end{tikzpicture}
 \end{minipage}
 \caption{An example graph (left) with the corresponding shortest path graph $H_A$ from $A$ (right).}\label{fig:SPGexample}
\end{figure}

%% file: graphic.tex
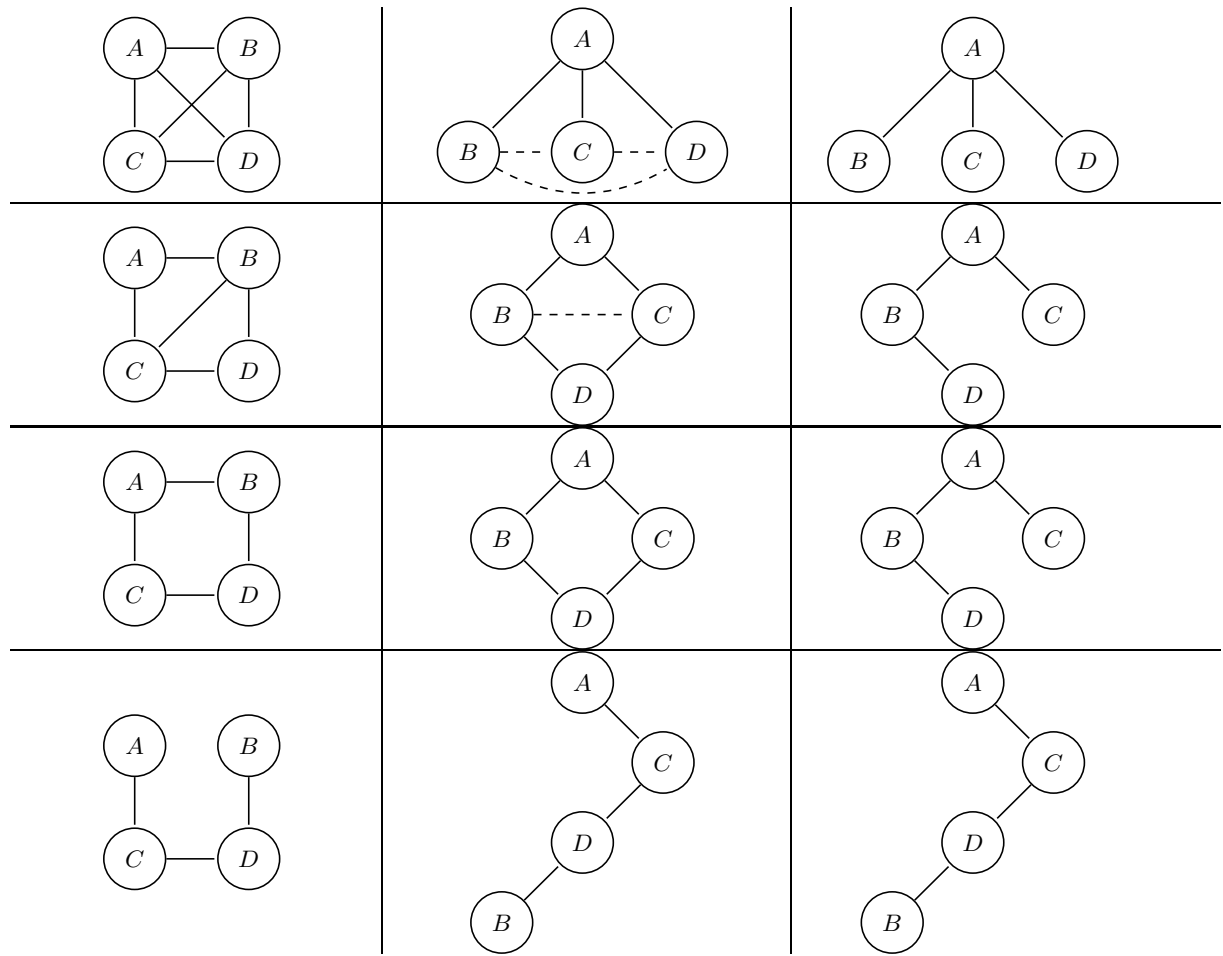
\begin{figure}[h]
 \begin{minipage}{0.3\textwidth}
 \centering
\begin{tikzpicture}[-,>=stealth',shorten >=1pt,auto,node distance=1.5cm,
                    semithick]
  \tikzstyle{every state}=[fill=none,draw=black,text=black]

  \node[state] (A)              {$A$};
  \node[state]         (B) [right of=A] {$B$};
  \node[state]         (C) [below of=A] {$C$};
  \node[state]         (D) [below of=B] {$D$};

  \path (A) edge              node {} (B)
            edge              node {} (C)
            edge              node {} (D)
        (B) edge              node {} (C)
	    edge              node {} (D)
        (C) edge              node {} (D);
\end{tikzpicture}
 \end{minipage}
\vrule
 \begin{minipage}{0.33\textwidth}
 \centering
  \begin{tikzpicture}[-,>=stealth',shorten >=1pt,auto,node distance=1.5cm,
                    semithick]
  \tikzstyle{every state}=[fill=none,draw=black,text=black]

  \node[state] (A)              {$A$};
  \node[state]         (B) [left of=C]  {$B$};
  \node[state]         (C) [below of=A] {$C$};
  \node[state]         (D) [right of=C] {$D$};

  \path (A) edge                      node {} (B)
            edge                      node {} (C)
            edge                      node {} (D)
	(B) edge [dashed]             node {} (C)
            edge [bend right, dashed] node {} (D)
	(C) edge [dashed]             node {} (D);
\end{tikzpicture}
 \end{minipage}
 \vrule
\begin{minipage}{0.3\textwidth}
\centering
  \begin{tikzpicture}[-,>=stealth',shorten >=1pt,auto,node distance=1.5cm,
                    semithick]
  \tikzstyle{every state}=[fill=none,draw=black,text=black]

  \node[state] (A)              {$A$};
  \node[state]         (B) [left of=C]  {$B$};
  \node[state]         (C) [below of=A] {$C$};
  \node[state]         (D) [right of=C] {$D$};

  \path (A) edge                      node {} (B)
            edge                      node {} (C)
            edge                      node {} (D);
\end{tikzpicture}
 \end{minipage}
\hrule
\begin{minipage}{0.3\textwidth}
\centering
\begin{tikzpicture}[-,>=stealth',shorten >=1pt,auto,node distance=1.5cm,
                    semithick]
  \tikzstyle{every state}=[fill=none,draw=black,text=black]

  \node[state] (A)              {$A$};
  \node[state]         (B) [right of=A] {$B$};
  \node[state]         (C) [below of=A] {$C$};
  \node[state]         (D) [below of=B] {$D$};

  \path (A) edge              node {} (B)
            edge              node {} (C)
        (B) edge              node {} (C)
	    edge              node {} (D)
        (C) edge              node {} (D);
\end{tikzpicture}
\end{minipage}
\vrule
 \begin{minipage}{0.33\textwidth}
 \centering
  \begin{tikzpicture}[-,>=stealth',shorten >=1pt,auto,node distance=1.5cm,
                    semithick]
  \tikzstyle{every state}=[fill=none,draw=black,text=black]

  \node[state] (A)              	      {$A$};
  \node[state]         (B) [below left of=A]  {$B$};
  \node[state]         (C) [below right of=A] {$C$};
  \node[state]         (D) [below right of=B] {$D$};

  \path (A) edge          node {} (B)
            edge          node {} (C)
	(B) edge [dashed] node {} (C)
            edge          node {} (D)
	(C) edge          node {} (D);
\end{tikzpicture}
 \end{minipage}
 \vrule
\begin{minipage}{0.3\textwidth}
\centering
  \begin{tikzpicture}[-,>=stealth',shorten >=1pt,auto,node distance=1.5cm,
                    semithick]
  \tikzstyle{every state}=[fill=none,draw=black,text=black]

  \node[state] (A)              	      {$A$};
  \node[state]         (B) [below left of=A]  {$B$};
  \node[state]         (C) [below right of=A] {$C$};
  \node[state]         (D) [below right of=B] {$D$};

  \path (A) edge          node {} (B)
            edge          node {} (C)
	(B) edge          node {} (D);
\end{tikzpicture}
 \end{minipage}
\hrule
 \begin{minipage}{0.3\textwidth}
 \centering
\begin{tikzpicture}[-,>=stealth',shorten >=1pt,auto,node distance=1.5cm,
                    semithick]
  \tikzstyle{every state}=[fill=none,draw=black,text=black]

  \node[state] (A)              {$A$};
  \node[state]         (B) [right of=A] {$B$};
  \node[state]         (C) [below of=A] {$C$};
  \node[state]         (D) [below of=B] {$D$};

  \path (A) edge              node {} (B)
            edge              node {} (C)
        (B) edge              node {} (D)
        (C) edge              node {} (D);
\end{tikzpicture}
\end{minipage}
\vrule
 \begin{minipage}{0.33\textwidth}
 \centering
  \begin{tikzpicture}[-,>=stealth',shorten >=1pt,auto,node distance=1.5cm,
                    semithick]
  \tikzstyle{every state}=[fill=none,draw=black,text=black]

  \node[state] (A)              	      {$A$};
  \node[state]         (B) [below left of=A]  {$B$};
  \node[state]         (C) [below right of=A] {$C$};
  \node[state]         (D) [below right of=B] {$D$};

  \path (A) edge          node {} (B)
            edge          node {} (C)
	(B) edge          node {} (D)
	(C) edge          node {} (D);
\end{tikzpicture}
 \end{minipage}
 \vrule
\begin{minipage}{0.3\textwidth}
\centering
  \begin{tikzpicture}[-,>=stealth',shorten >=1pt,auto,node distance=1.5cm,
                    semithick]
  \tikzstyle{every state}=[fill=none,draw=black,text=black]

  \node[state] (A)              	      {$A$};
  \node[state]         (B) [below left of=A]  {$B$};
  \node[state]         (C) [below right of=A] {$C$};
  \node[state]         (D) [below right of=B] {$D$};

  \path (A) edge          node {} (B)
            edge          node {} (C)
	(B) edge          node {} (D);
\end{tikzpicture}
 \end{minipage}
\hrule
 \begin{minipage}{0.3\textwidth}
 \centering
\begin{tikzpicture}[-,>=stealth',shorten >=1pt,auto,node distance=1.5cm,
                    semithick]
  \tikzstyle{every state}=[fill=none,draw=black,text=black]

  \node[state] (A)              {$A$};
  \node[state]         (B) [right of=A] {$B$};
  \node[state]         (C) [below of=A] {$C$};
  \node[state]         (D) [below of=B] {$D$};

  \path (A) edge              node {} (C)
        (B) edge              node {} (D)
        (C) edge              node {} (D);
\end{tikzpicture}
\end{minipage}
\vrule
 \begin{minipage}{0.33\textwidth}
 \centering
  \begin{tikzpicture}[-,>=stealth',shorten >=1pt,auto,node distance=1.5cm,
                    semithick]
  \tikzstyle{every state}=[fill=none,draw=black,text=black]

  \node[state] (A)              	      {$A$};
  \node[state]         (C) [below right of=A] {$C$};
  \node[state]         (D) [below left of=C] {$D$};
    \node[state]         (B) [below left of=D]  {$B$};

  \path (A) edge          node {} (C)
	(B) edge          node {} (D)
	(C) edge          node {} (D);
\end{tikzpicture}
 \end{minipage}
 \vrule
\begin{minipage}{0.3\textwidth}
\centering
  \begin{tikzpicture}[-,>=stealth',shorten >=1pt,auto,node distance=1.5cm,
                    semithick]
  \tikzstyle{every state}=[fill=none,draw=black,text=black]

  \node[state] (A)              	      {$A$};
  \node[state]         (C) [below right of=A] {$C$};
  \node[state]         (D) [below left of=C] {$D$};
    \node[state]         (B) [below left of=D]  {$B$};

  \path (A) edge          node {} (C)
	(B) edge          node {} (D)
	(C) edge          node {} (D);
\end{tikzpicture}
 \end{minipage}
 \caption{An example of the data structures under the following series of deletions: $(A,D),(B,C),(A,B)$. The left hand column represents the original graph $G$, the middle column is the data structure $H$, and the right hand column is the data structure $T$ which utilises the reduced memory technique.}\label{fig:ESexample}
\end{figure}

%% file: graphic-falling.tex
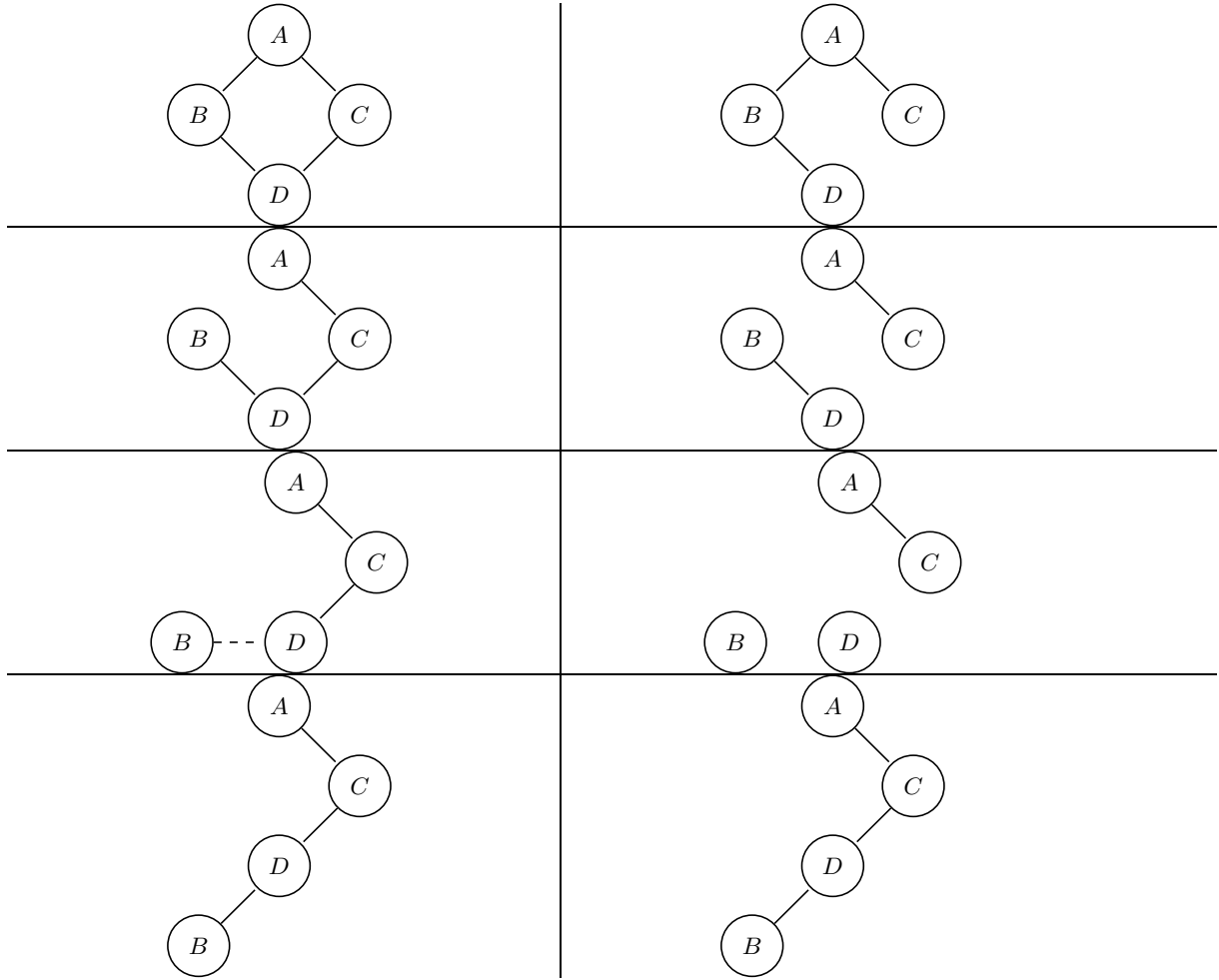
\begin{figure}[h]
 \begin{minipage}{0.45\textwidth}
 \centering
  \begin{tikzpicture}[-,>=stealth',shorten >=1pt,auto,node distance=1.5cm,
                    semithick]
  \tikzstyle{every state}=[fill=none,draw=black,text=black]

  \node[state] (A)              	      {$A$};
  \node[state]         (B) [below left of=A]  {$B$};
  \node[state]         (C) [below right of=A] {$C$};
  \node[state]         (D) [below right of=B] {$D$};

  \path (A) edge          node {} (B)
            edge          node {} (C)
	(B) edge          node {} (D)
	(C) edge          node {} (D);
\end{tikzpicture}
 \end{minipage}
 \vrule
\begin{minipage}{0.45\textwidth}
\centering
  \begin{tikzpicture}[-,>=stealth',shorten >=1pt,auto,node distance=1.5cm,
                    semithick]
  \tikzstyle{every state}=[fill=none,draw=black,text=black]

  \node[state] (A)              	      {$A$};
  \node[state]         (B) [below left of=A]  {$B$};
  \node[state]         (C) [below right of=A] {$C$};
  \node[state]         (D) [below right of=B] {$D$};

  \path (A) edge          node {} (B)
            edge          node {} (C)
	(B) edge          node {} (D);
\end{tikzpicture}
 \end{minipage}
\hrule

\begin{minipage}{0.45\textwidth}
 \centering
  \begin{tikzpicture}[-,>=stealth',shorten >=1pt,auto,node distance=1.5cm,
                    semithick]
  \tikzstyle{every state}=[fill=none,draw=black,text=black]

  \node[state] (A)              	      {$A$};
  \node[state]         (B) [below left of=A]  {$B$};
  \node[state]         (C) [below right of=A] {$C$};
  \node[state]         (D) [below right of=B] {$D$};

  \path (A) edge          node {} (C)
	(B) edge          node {} (D)
	(C) edge          node {} (D);
\end{tikzpicture}
 \end{minipage}
 \vrule
\begin{minipage}{0.45\textwidth}
\centering
  \begin{tikzpicture}[-,>=stealth',shorten >=1pt,auto,node distance=1.5cm,
                    semithick]
  \tikzstyle{every state}=[fill=none,draw=black,text=black]

  \node[state] (A)              	      {$A$};
  \node[state]         (B) [below left of=A]  {$B$};
  \node[state]         (C) [below right of=A] {$C$};
  \node[state]         (D) [below right of=B] {$D$};

  \path (A) edge          node {} (C)
	(B) edge          node {} (D);
\end{tikzpicture}
 \end{minipage}
\hrule

\begin{minipage}{0.45\textwidth}
 \centering
  \begin{tikzpicture}[-,>=stealth',shorten >=1pt,auto,node distance=1.5cm,
                    semithick]
  \tikzstyle{every state}=[fill=none,draw=black,text=black]

  \node[state] (A)              	      {$A$};
  \node[state]         (C) [below right of=A] {$C$};
  \node[state]         (D) [below left of=C] {$D$};
  \node[state]         (B) [left of=D]  {$B$};

  \path (A) edge          node {} (C)
	(B) edge[dashed]  node {} (D)
	(C) edge          node {} (D);
\end{tikzpicture}
 \end{minipage}
 \vrule
\begin{minipage}{0.45\textwidth}
\centering
  \begin{tikzpicture}[-,>=stealth',shorten >=1pt,auto,node distance=1.5cm,
                    semithick]
  \tikzstyle{every state}=[fill=none,draw=black,text=black]

  \node[state] (A)              	      {$A$};
  \node[state]         (C) [below right of=A] {$C$};
  \node[state]         (D) [below left of=C] {$D$};
  \node[state]         (B) [left of=D]  {$B$};

  \path (A) edge          node {} (C);
\end{tikzpicture}
 \end{minipage}
\hrule

 \begin{minipage}{0.45\textwidth}
 \centering
  \begin{tikzpicture}[-,>=stealth',shorten >=1pt,auto,node distance=1.5cm,
                    semithick]
  \tikzstyle{every state}=[fill=none,draw=black,text=black]

  \node[state] (A)              	      {$A$};
  \node[state]         (C) [below right of=A] {$C$};
  \node[state]         (D) [below left of=C] {$D$};
    \node[state]         (B) [below left of=D]  {$B$};

  \path (A) edge          node {} (C)
	(B) edge          node {} (D)
	(C) edge          node {} (D);
\end{tikzpicture}
 \end{minipage}
 \vrule
\begin{minipage}{0.45\textwidth}
\centering
  \begin{tikzpicture}[-,>=stealth',shorten >=1pt,auto,node distance=1.5cm,
                    semithick]
  \tikzstyle{every state}=[fill=none,draw=black,text=black]

  \node[state] (A)              	      {$A$};
  \node[state]         (C) [below right of=A] {$C$};
  \node[state]         (D) [below left of=C] {$D$};
    \node[state]         (B) [below left of=D]  {$B$};

  \path (A) edge          node {} (C)
	(B) edge          node {} (D)
	(C) edge          node {} (D);
\end{tikzpicture}
 \end{minipage}
 \caption{An example of the steps taken by the algorithm when the edge $(A,B)$ is deleted from the example graph. The left hand column representsthe data structure $H$ and the right hand column is the data structure $T$ which utilises the reduced memory technique.}\label{fig:ESfalling}
\end{figure}